\documentclass[pra, reprint, amsmath, amssymb, aps,superscriptaddress]{revtex4-2}

\usepackage{mathtools}
\usepackage{tabularx}
\usepackage{booktabs}
\usepackage{xcolor}
\usepackage{siunitx}

\newcommand{\eqnref}[1]{Eq.~\eqref{#1}}
\newcommand{\figref}[1]{Fig.~\ref{#1}}

\newcommand{\appref}[1]{App.~\ref{#1}}

\usepackage{hyperref}
\hypersetup{colorlinks=true, citecolor=blue, urlcolor=blue, linkcolor=blue}

\begin{document}

\title{Optimal Position Detection of an Optically Levitated Mie Particle}

\author{Long Wang}
\affiliation{CAS Key Laboratory of Quantum Information, University of Science and Technology of China, Hefei 230026, China}
\affiliation{CAS Center for Excellence in Quantum Information and Quantum Physics, University of Science and Technology of China, Hefei 230026, China}

\author{Lei-Ming Zhou}
\affiliation{Department of Optical Engineering, School of Physics, Hefei University of Technology, Hefei 230601, China}

\author{Yuan Tian}
\author{Lyu-Hang Liu}
\author{Guang-Can Guo}
\author{Yu Zheng}
\email{bigz@ustc.edu.cn}
\author{Fang-Wen Sun}
\affiliation{CAS Key Laboratory of Quantum Information, University of Science and Technology of China, Hefei 230026, China}
\affiliation{CAS Center for Excellence in Quantum Information and Quantum Physics, University of Science and Technology of China, Hefei 230026, China}

\date{\today}

\begin{abstract}
  We theoretically investigate the problem of position detection of an optically levitated Mie particle.
  The information radiation field (IRF) is proposed and defined to characterize the scattered light carrying complete information about the center-of-mass (c.m.) motion of the particle.
  Based on the IRF, we suggest an optimal detection scheme for the position of arbitrary particles.
  We calculate both the information losses of objective collection and mode-matching in levitated optomechanical experiments.
  Our results conclude that the backward detection scheme, using an incident Gaussian beam focused by a high numerical aperture lens, provides sufficient information to achieve the quantum ground state through cooling of the three-dimensional c.m. motion of the Mie particle.
\end{abstract}

\maketitle

\section*{Introduction}
Levitated optomechanics offers a promising platform for studying macroscopic quantum phenomena, and has demonstrated remarkable achievements \cite{gonzalez-ballestero2021levitodynamics,millen2020optomechanics,winstone2023levitated}.
Particularly, significant progress has been made experimentally in preparing the quantum ground state of the c.m. motion of a nanoparticle \cite{delic2020cooling,magrini2021realtime,tebbenjohanns2021quantum,kamba2022optical,ranfagni2022twodimensional,piotrowski2023simultaneous,kamba2023revealing}.
These experiments, which rely on optically detecting particle motion, are limited by the Heisenberg uncertainty principle \cite{clerk2010introduction,bowen2015quantum}.
In the canonical optical detection setup, the phase of the scattered light encoding the information of the particle's position \cite{gittes1998interference,jones2015optical,zheng2019threedimensional} is obtained by the homodyne detection.
Concurrently, fluctuations in the radiation pressure induced by the scattered light inevitably disturb the particle's motion, limiting the position measurement rate $\Gamma_{\mathrm{meas}}$, i.e. $\Gamma_{\mathrm{meas}} \le \Gamma_{\mathrm{ba}}$ \cite{clerk2010introduction,bowen2015quantum}.
This upper bound, $\Gamma_{\mathrm{ba}}$, is the measurement backaction rate,  also known as the photon recoil heating rate \cite{jain2016direct}.
An optimal position detection scheme (where $\Gamma_{\mathrm{meas}} = \Gamma_{\mathrm{ba}}$) for an isotropic dipolar scatterer has been proposed under the Rayleigh approximation \cite{tebbenjohanns2019optimal}.
This scheme has been used to achieve the quantum ground state cooling of the motion of a nanoparticle ($10^{9}$ atomic mass units) along the light propagation $z$-axis \cite{magrini2021realtime, tebbenjohanns2021quantum}.
However, the quantum ground state cooling of the c.m. motion of particles beyond Rayleigh approximation has not been achieved.
Bringing particles with larger masses into the quantum regime  \cite{neumeier2024fast,riera-campeny2024wigner,roda-llordes2024macroscopic,roda-llordes2024numerical,romero-isart2011large,romero-isart2011optically,romero-isart2011quantum} is crucial for advancing frontier physics, including tests of quantum gravity \cite{aspelmeyer2014cavity,kaltenbaek2023research}, gravitational wave detection \cite{arvanitaki2013detecting,aggarwal2022searching,direkci2024macroscopic} and dark matter detection \cite{afek2021limits,afek2022coherent,carney2023searches,kawasaki2020high,monteiro2020search,moore2021searching,priel2022dipole,priel2022search,rademacher2020quantum}.

Continuous position measurement of a mechanical oscillator is governed by the Heisenberg uncertainty principle \cite{clerk2010introduction,bowen2015quantum, wilson2015measurementbased}, which limits the position detection efficiency $\eta = \Gamma_{\mathrm{meas}} / \Gamma_{\mathrm{ba}}$ to $\eta \le 1$.
Experimental observations have shown that achieving quantum ground state cooling through active feedback \cite{magrini2021realtime,tebbenjohanns2021quantum,rossi2018measurementbased,wilson2015measurementbased,huang2024roomtemperature} is constrained by this position detection efficiency.
Specifically, the minimum achievable mean phonon-occupation number of the oscillator is given by $\bar{n}_{\min} = (\eta^{-1 /2} - 1) /2$.
Therefore, improving the position detection efficiency is crucial for generating the quantum state of a particle.
In the levitated optomechanics, various methods have been proposed to enhance collection efficiency of a nanoparticle.
These methods include collecting laterally scattered photons by deep parabolic mirrors \cite{alber2017focusing} or by self-homodyne detection \cite{cerchiari2021position,dania2022position}, suppressing lateral photon recoil by a hollow hemispherical mirror \cite{weiser2024back} and shining squeezed light \cite{gonzalez-ballestero2023suppressing}.
Additionally, studies \cite{pflanzer2012masterequation, seberson2020distribution, maurer2023quantum} have investigated the interaction between light and a Mie particle.
However, the optimal position detection scheme for a Mie particle, where the particle size is comparable to or larger than the light wavelength $\lambda_0$, remains unclear.
Further research is needed to effectively extract information about the c.m. motion of Mie particles from the total scattering light.

In this paper, we theoretically analyze the problem of optimally measuring the position of a Mie particle in a focused light field.
We introduce the IRF as the differential of the scattering field with respect to the particle's displacement to characterize the information photons coupling to the three-dimensional center-of-mass (c.m.) motion of the particle.
To detect particle motion at the Heisenberg limit and achieve optimal detection efficiency $\eta = 1$, it is necessary to consider homodyne detection of the IRF.
We evaluate both the information collection efficiency and the mode-matching efficiency in typical levitated optomechanical configurations.
Our findings indicate that the backward detection scheme, using an incident beam with a high numerical aperture ($\mathrm{NA}_{\mathrm{in}} = 0.9$), can potentially achieve three-dimensional c.m. ground-state cooling of a Mie particle with masses ranging from $10^{9}$ to $10^{14}$ atomic mass units.
This scheme can attain information collection efficiencies exceeding $0.8$ for all three axes.
Moreover, the total detection efficiencies of $0.8$ for the longitudinal $z$-axis and $0.4$ for the lateral $x, y$-axes can be achieved using fiber-based balanced homodyne detection.

\section{Theory fundamentals of ideal position measurement}
The theory of optimal position detection for nanoparticles and dipoles has been proposed and discussed in previous works \cite{tebbenjohanns2019optimal}.
In this section, we expand the theory of optimal position detection to arbitrary particles, including Mie-sized particles.

\begin{figure}
  \centering
  \includegraphics[width=\linewidth]{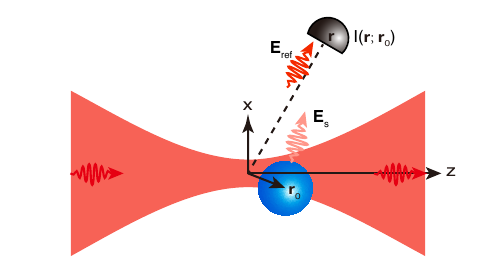}
  \caption{Sketch of the ideal position detection scheme. A dielectric sphere is displaced $\mathbf{r}_{0}$ relative to the origin, which is defined as the focal point of the incident beam from left to right. A reference field $\mathbf{E}_{\mathrm{ref}}(\mathbf{r})$ is added to homodyne detection of the phase of the scattering field $\mathbf{E}_{\mathrm{s}}(\mathbf{r}; \mathbf{r}_{0})$ at a point $\mathbf{r}$ in the far field. The intensity distribution $I(\mathbf{r}; \mathbf{r}_{0})$ is detected by an array of detectors in the far field.}
  \label{fig:sketch}
\end{figure}

We consider an $x$-polarized Gaussian beam propagating along the $z$-axis, which is strongly focused, as illustrated in \figref{fig:sketch}.
The particle is assumed to be trapped at the focal point of the objective lens.
The origin of the coordinate system is defined by this focus, with the particle undergoing minute motions around the origin.
%

The information about the particle c.m. motion is primarily encoded in the phase of the scattered light.
To extract the phase and to amplify the signal, the scattered field $\mathbf{E}_{\mathrm{s}}$ is interfered with a reference field $\mathbf{E}_{\mathrm{ref}}$ in the homodyne detection.
The intensity of the reference light must be significantly greater than the scattered light, i.e. $|\mathbf{E}_{\mathrm{ref}}| \gg |\mathbf{E}_{\mathrm{s}}|$.
Considering a small displacement $\mathbf{r}_{0}$ of the particle, the intensity distribution at the detector at point $\mathbf{r}$ in the far field is
\begin{equation}\label{eq:I_int}
  \begin{aligned}
    I(\mathbf{r}; \mathbf{r}_{0}) & \propto |\mathbf{E}_{\mathrm{tot}}(\mathbf{r}; \mathbf{r}_{0})|^{2} = |\mathbf{E}_{\mathrm{ref}}(\mathbf{r}) + \mathbf{E}_{\mathrm{s}}(\mathbf{r}; \mathbf{r}_{0})|^{2}                                                       \\
                                  & = |\mathbf{E}_{\mathrm{ref}}(\mathbf{r})|^{2} + 2 \mathrm{Re}\{ \mathbf{E}_{\mathrm{ref}}^{*}(\mathbf{r}) \mathbf{E}_{\mathrm{s}}(\mathbf{r}; \mathbf{r}_{0}) \} + |\mathbf{E}_{\mathrm{s}}(\mathbf{r}; \mathbf{r}_{0})|^{2},
  \end{aligned}
\end{equation}
where the first term is a constant offset and the second term is the signal to detect.
In our case, the intensity of the reference light is significantly greater than that of the scattered light, $|\mathbf{E}_{\mathrm{ref}}| \gg |\mathbf{E}_{\mathrm{s}}|$.
Then the third term of \eqnref{eq:I_int} is small compared to the other two and can be neglected.
Thus it is needed to analyze the second term to propose the optimal position detection protocol.

\subsection{Information radiation field}
Mathematically, the scattered field can be expanded to first order in the small displacement limit as
\begin{equation}\label{eq:expand_Es}
  \mathbf{E}_{\mathrm{s}}(\mathbf{r};r_{0\mu}) = \mathbf{E}_{\mathrm{s}}(\mathbf{r};0) + r_{0\mu} \left. \frac{\partial \mathbf{E}_{\mathrm{s}}(\mathbf{r};r_{0\mu})}{\partial r_{0\mu}} \right|_{r_{0\mu} = 0},
\end{equation}
where the displacement $r_{0\mu}$ of the particle along $\mu$-axis ($\mu = x, y, z$) is much smaller than the incident light wavelength $\lambda_0$.
This can be guaranteed by high detection bandwidth and feedback cooling.
The zero-order expansion term characterizes the elastically scattered photons, which carry no position information.
In contrast, the first-order expansion term is defined as the Information Radiation Field (IRF), expressed as
\begin{equation}\label{eq:def_IRF}
  \mathbf{E}_{\mu}(\mathbf{r}) \equiv r_{0\mu} \left. \frac{\partial \mathbf{E}_{\mathrm{s}}(\mathbf{r};r_{0\mu})}{\partial r_{0\mu}} \right|_{r_{0\mu} = 0},
\end{equation}
which characterizes the scattered photons carrying complete information about the particle's position.
This implies that the information is not uniformly distributed within the scattered light \cite{tebbenjohanns2019optimal, maurer2023quantum} and cannot be explicitly expressed by the scattered field $\mathbf{E}_{\mathrm{s}}$.
In other words, only a fraction of scattered photons contain position information, while the majority undergo elastic scattering and carry no information.
Specifically, the ratio of the inelastically scattered photons, referred to as information photons, is approximately $10^{-12}$ for a wavelength $\lambda_0 = \SI{1064}{\nm}$ and a displacement $r_{0\mu} = \SI{1}{\pm}$.
It is helpful to rewrite the \eqnref{eq:I_int} as
\begin{equation}\label{eq:I_info}
  \begin{aligned}
    I(\mathbf{r}; \mathbf{r}_{0}) & \propto |\mathbf{E}_{\mathrm{tot}}(\mathbf{r};0)|^{2} + 2 \mathrm{Re}\{ \mathbf{E}_{\mathrm{tot}}^{*}(\mathbf{r};0) \mathbf{E}_{\mu}(\mathbf{r}) \} + |\mathbf{E}_{\mu}(\mathbf{r})|^{2} \\
                                  & = |\mathbf{E}_{\mathrm{tot}}(\mathbf{r};0)|^{2} + 2 \mathrm{Re}\{ \mathbf{E}_{\mathrm{ref}}^{*}(\mathbf{r}) \mathbf{E}_{\mu}(\mathbf{r}) \} + |\mathbf{E}_{\mu}(\mathbf{r})|^{2},
  \end{aligned}
\end{equation}
where the first term $\mathbf{E}_{\mathrm{tot}}(\mathbf{r};0) = \mathbf{E}_{\mathrm{ref}}(\mathbf{r}) + \mathbf{E}_{\mathrm{s}}(\mathbf{r}; 0)$ contributes the total constant offset as well as the predominant shot noise.
Particularly, the elastically scattered field $\mathbf{E}_{\mathrm{s}}(\mathbf{r}; \mathbf{r}_{0\mu})$ only provides the shot noise without providing positional information or interfering effects.
The normalized IRF $\mathbf{F}_{\mu}$ is defined as
\begin{equation}\label{eq:normalized_IRF}
  \mathbf{F}_{\mu}(\mathbf{r}) \equiv \frac{ \mathbf{E}_{\mu}(\mathbf{r})}{\sqrt{\int_{4\pi} |\mathbf{E}_{\mu}(\mathbf{r})|r^{2} \mathrm{d}\Omega}}.
\end{equation}
%

\subsection{Optimal position detection}
When all light signals within a 4$\pi$ solid angle are collected and the constant offset $|\mathbf{E}_{\mathrm{tot}}(\mathbf{r};0)|^{2}$ is cancelled by a balanced detector, we obtain a power signal proportional to the displacement as
\begin{equation}\label{eq:cali}
  P_{\mu} = C_{\mu} r_{0\mu}= \frac{1}{2} c \epsilon_0 \left| \int_{4\pi} 2 \mathrm{Re}\left\{ \mathbf{E}_{\mathrm{ref}}^{*}(\mathbf{r}) \mathbf{E}_{\mu}(\mathbf{r}) \right\} r^{2} \mathrm{d}\Omega \right|.
\end{equation}
where $\Omega$ is the solid angle of the detector and the coefficient $C_{\mu}$ converting the displacement to the power signal for $\mu = x, y, z$-axis motion.
The balanced detection can eliminate the constant offset of $|\mathbf{E}_{\mathrm{tot}}(\mathbf{r})|^{2}$; however, it does not eliminate the shot noise, which also leads to a shot noise in the position signal or position imprecision noise.
The shot noise of light field is characterized by a power spectral density (PSD) of $S_{PP} = \hbar \omega_0 P_{\mathrm{det}}$, where the detected light power $P_{\mathrm{det}}$ is dominated by the reference filed $\mathbf{E}_{\mathrm{ref}}$.
This leads to the spectral density $S_{r_{\mu} r_{\mu}}^{\mathrm{imp}}$ of the position imprecision noise
\begin{equation}\label{eq:S_imp}
  S_{r_{\mu} r_{\mu}}^{\mathrm{imp}}\equiv C_{\mu}^{-2} S_{PP} = \frac{2 \hbar \omega_0}{c \epsilon_0} \frac{\int_{4\pi} |\mathbf{E}_{\mathrm{ref}}|^{2} r^{2} \mathrm{d}\Omega}{\left| \int_{4\pi} 2 \mathrm{Re}\left\{ \mathbf{E}_{\mathrm{ref}}^{*} \mathbf{E}_{\mu} \right\} r^{2} \mathrm{d}\Omega \right|^{2} r_{0\mu}^{-2}}.
\end{equation}
The precision of a noisy measurement improves with measurement duration.
However, a quantum-limited measurement inevitably disrupts the system, thereby restricting the duration for which one can measure a quantum state before its complete destruction by the measurement process \cite{clerk2010introduction}.
The measurement rate $\Gamma_{\mu}^{\mathrm{meas}}$ is defined as the rate where our measurement is able to distinguish the displacement equivalent to the zero-point position fluctuation:
\begin{equation}\label{eq:Gamma_meas}
  \Gamma_{\mu}^{\mathrm{meas}} \equiv \frac{r_{0\mu}^{2}}{4 S_{r_{\mu} r_{\mu}}^{\mathrm{imp}}}   = \frac{c \epsilon_0}{2 \hbar \omega_0} \frac{|\int_{4\pi} \mathrm{Re}\{ \mathbf{E}_{\mathrm{ref}}^{*} \cdot \mathbf{E}_{\mu} \} r^{2} \mathrm{d}\Omega|^{2}}{\int _{4\pi}|\mathbf{E}_{\mathrm{ref}}|^{2} r^{2} \mathrm{d}\Omega},
\end{equation}
where the displacement $r_{0\mu}$ is specifically considered as the zero-point position fluctuation of a mechanical oscillator (i.e. $r_{0\mu} = \sqrt{\hbar / (2 m \Omega)}$, where $m$ and $\Omega_{\mu}$ are the mass and mechanical frequency of the particle, respectively) as well as $r_{0\mu}$ in \eqnref{eq:def_IRF}.
According to the Cauchy-Schwarz inequality, the measurement rate $\Gamma_{\mu}^{\mathrm{meas}}$ have an upper bound, the so-called photon recoil heating rate $\Gamma_{\mu}^{\mathrm{ba}}$ \cite{jain2016direct}, namely
\begin{equation}\label{eq:C-S_ineq}
  \Gamma_{\mu}^{\mathrm{meas}} \le \Gamma_{\mu}^{\mathrm{ba}} = \frac{c \epsilon_0}{2 \hbar \omega_0} \int |\mathbf{E}_{\mu}(\mathbf{r})|^{2} r^{2} \mathrm{d}\Omega.
\end{equation}
This is consistent with the definition of Refs. \cite{pflanzer2012masterequation,seberson2020distribution,maurer2023quantum}.
From the perspective of quantum theory \cite{maurer2023quantum}, the IRF describes the scattered field resulting from the incident beam by absorption or excitation of a mechanical phonon when considering the displacement $r_{0\mu}$ as the zero-point motion $\sqrt{\hbar /(2 m \Omega_{\mu})}$.
Furthermore, our IRF $\mathbf{E}_{\mu}$ can be analytically expressed as a sum over discrete angular momentum and polarization indices $p, l, m$ by applying the Fourier transform to the transform amplitude $\tau_{\kappa \mu}$ in Ref. \cite{maurer2023quantum}.

Finally, the total position detection efficiency in the interferometric measurement process is defined as
\begin{equation}\label{eq:def_eta}
  \eta_{\mu} \equiv \frac{\Gamma_{\mu}^{\mathrm{meas}}}{\Gamma_{\mu}^{\mathrm{ba}}} = \frac{|\int_{4\pi} \mathrm{Re}\{ \mathbf{E}_{\mathrm{ref}}^{*} \cdot \mathbf{E}_{\mu} \} r^{2} \mathrm{d}\Omega|^{2}}{\int _{4\pi}|\mathbf{E}_{\mathrm{ref}}|^{2} r^{2} \mathrm{d}\Omega \int_{4\pi} |\mathbf{E}_{\mu}|^{2} r^{2} \mathrm{d}\Omega} \le 1.
\end{equation}
To achieve optimal position detection, i.e. position detection efficiency $\eta_{\mu} = 1$,  three conditions must be met:
(i) ensuring that the intensity of the reference field is significantly greater than that of the scattering field $\mathbf{E}_{\mathrm{ref}} \gg \mathbf{E}_{s}$,
(ii) implementing a mode-matching reference field $\mathbf{E}_{\mathrm{ref}} = \lambda \mathbf{E}_{\mu}$ where $\lambda$ is a real number,
and (iii) collecting all light information in the full $4\pi$ solid angle.
\section{Detection efficiency in realistic position measurement}
However, realistic measurements are constrained by two factors \cite{tebbenjohanns2019optimal,magrini2021realtime}:
(i) the information collection efficiency limited by the numerical aperture of the collection lens
and (ii) mode-matching efficiency between the reference field and the IRF in a photodetector.
Therefore, it is crucial to investigate both the information collection efficiency $\eta_{\mu}^{\mathrm{c}}$ and the mode-matching efficiency $\eta_{\mu}^{\mathrm{m}}$ in the measurement progresses.

\subsection{Information collection efficiency and mode-matching efficiency}
Firstly, the information collection efficiency $\eta_{\mu}^{\mathrm{c}}$ is defined as the ratio of information collected by an objective lens for a limited collection angle $\Omega_{\mathrm{c}}$ over the full $4\pi$ space, expressed as
\begin{equation}\label{eq:etac}
  \eta_{\mu}^{\mathrm{c}} \equiv \lim_{r \to \infty} \frac{\int_{\Omega_{\mathrm{c}}} |\mathbf{E}_{\mu}(\mathbf{r})|^{2} r^{2} \mathrm{d}\Omega}{\int_{4\pi} |\mathbf{E}_{\mu}(\mathbf{r})|^{2} r^{2} \mathrm{d}\Omega}.
\end{equation}

Then the mode-matching efficiency $\eta_{\mu}^{\mathrm{m}}$ in the measurement process is defined as \cite{vamivakas2007phasesensitive,magrini2021realtime}:
\begin{equation}\label{eq:etam}
  \eta_{\mu}^{\mathrm{m}} \equiv \frac{|\mathrm{Re}\{\int \mathbf{E}^{*}_{\mathrm{ref}} \cdot \mathbf{E}_{\mu} \mathrm{d}X \mathrm{d}Y \}|^{2}}{\int |\mathbf{E}_{\mathrm{ref}}|^{2} \mathrm{d}X \mathrm{d}Y \int |\mathbf{E}_{\mu}|^{2} \mathrm{d}X \mathrm{d}Y}.
\end{equation}
This calculation involves integrating the fields over the $(X, Y)$ of the collection lens (refer to \appref{app:Field_cal} for details about field calculations).

Finally, the total position detection efficiency $\eta_{\mu}$
\footnote{Here we neglect other information losses including dissipation of thermal system (i.e., air molecular collision) and other photons losses in the measurement.
  This has been discussed extensively discussed in Ref. \cite{magrini2021realtime}.}
in \eqnref{eq:def_eta} can be expressed as the product of the information collection efficiency $\eta_{\mu}^{\mathrm{c}}$ and the mode-matching efficiency $\eta_{\mu}^{\mathrm{m}}$:
\begin{equation}\label{eq:eta}
  \eta_{\mu} = \eta_{\mu}^{\mathrm{c}} \cdot \eta_{\mu}^{\mathrm{m}} = \frac{|\mathrm{Re}\{\int \mathbf{E}^{*}_{\mathrm{ref}} \cdot \mathbf{E}_{\mu} \mathrm{d}X \mathrm{d}Y \}|^{2}}{\int |\mathbf{E}_{\mathrm{ref}}|^{2} \mathrm{d}X \mathrm{d}Y \int_{4\pi} |\mathbf{E}_{\mu}(\mathbf{r})|^{2} r^{2} \mathrm{d}\Omega},
\end{equation}
which is consistent with the definition of \eqnref{eq:def_eta}.
\subsection{Information collection}
\begin{figure*}[htbp]
  \centering
  \includegraphics[scale =1]{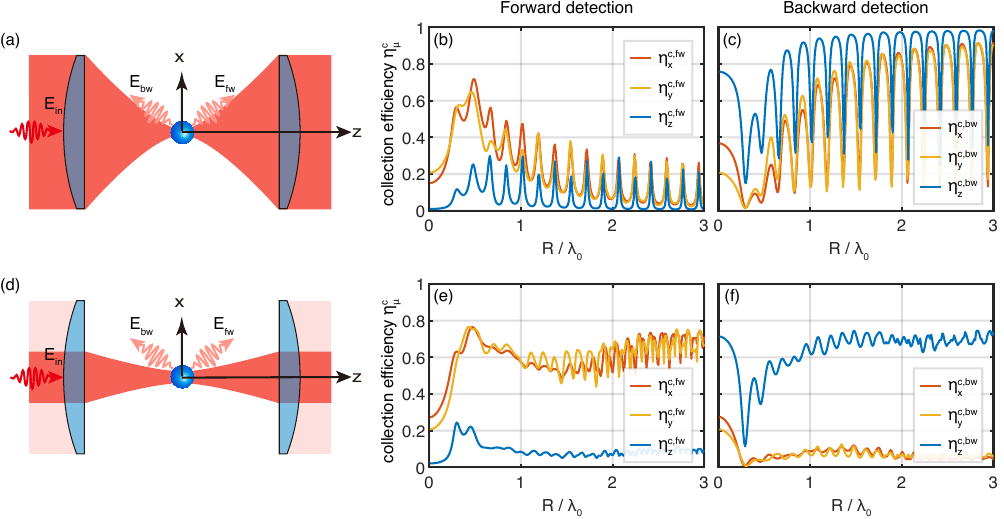}
  \caption{(a) Detection scheme of $\mathrm{NA}_{\mathrm{in}} = 0.9$.
  Here, $\mathbf{E}_{\mathrm{in}}$, $\mathbf{E}_{\mathrm{bw}}$ and $\mathbf{E}_{\mathrm{tw}}$ represent the incident field, backward and forward scattering fields, respectively.
  Both two objective lenses have a same numerical aperture $\mathrm{NA} = 0.9$.
  Information collection efficiency of backward detection $\eta_{\mu}^{\mathrm{c,bw}}$ (b) and of forward detection $\eta_{\mu}^{\mathrm{c,fw}}$ (c) for $\mu = x, y, z$-axis c.m. motion as a function of $R / \lambda_0$ in the configuration with $\mathrm{NA}_{\mathrm{in}} = 0.9$.
  (d) Detection scheme of $\mathrm{NA}_{\mathrm{in}} = 0.1$.
  Information collection efficiency of backward detection $\eta_{\mu}^{\mathrm{c,bw}}$ (e) and of forward detection $\eta_{\mu}^{\mathrm{c,fw}}$ (f) in the configuration with $\mathrm{NA}_{\mathrm{in}} = 0.1$.
  }
  \label{fig:detection_scheme}
\end{figure*}
%


First, we investigate the information collection efficiency in two configurations for Mie particles: one with a broad incident beam (\figref{fig:detection_scheme}(a)) and another with a narrowing incident beam within an effective numerical aperture of $\mathrm{NA}_{\mathrm{in}} = 0.1$ (\figref{fig:detection_scheme}(d)).
These two configurations correspond to the forward and backward detection, respectively.
Consequently, four detection schemes are illustrated in \figref{fig:detection_scheme}(b, c, e, f).
In this paper, the default physical parameters are detailed in Tab.~\ref{tab:parameters}.
\begin{table}
  \begin{tabularx}{0.9\columnwidth}{l r}
    \toprule
    \textbf{Parameters}              & \textbf{Description}         \\
    \midrule
    \midrule
    $n = 1.44$                       & refractive index (silica)    \\
    $\rho = 2200$ \unit{kg/m^3}      & mass density (silica)        \\
    $\lambda_0 = \SI{1064}{\nm}$     & laser wavelength             \\
    $P_{\mathrm{in}} = \SI{0.1}{\W}$ & incident beam power          \\
    $w_0 = \SI{2.25}{\mm}$           & waist radius                 \\
    $\mathrm{NA} = 0.9$              & objective numerical aperture \\
    $f = \SI{2}{\mm}$                & objective focal length       \\
    $R_{\mathrm{o}} = \SI{1.8}{\mm}$ & objective aperture radius    \\
    \bottomrule
  \end{tabularx}
  \caption{Default parameters considered throughout this article.}
  \label{tab:parameters}
\end{table}

The scattering of Mie particles ($R / \lambda_0 \gtrsim 0.5$) provides the high information collection efficiency in both forward and backward detection.
In particular, the configuration with $\mathrm{NA}_{\mathrm{in}} = 0.9$ in the backward detection shows remarkably the high information collection efficiency ($\eta_{\mu}^{\mathrm{c}} \approx 0.9$), for Mie particles with specific size parameters $R /\lambda_0$, as illustrated in \figref{fig:detection_scheme}(c).
The information collection efficiency demonstrates a periodic oscillation with $R / \lambda_0$ where the period is approximately $1 / 4n$, with $n$ being the refractive index of the particle.

The configuration with $\mathrm{NA}_{\mathrm{in}} = 0.1$ consistently exhibits high information collection efficiencies, regardless of $R / \lambda_0$, for lateral ($x,y$-axes) position in the forward detection \figref{fig:detection_scheme}(e) and longitudinal ($z$-axis) position in the backward detection \figref{fig:detection_scheme}(f).
However, with this slightly focused incident beam (close to a plane wave), the Mie resonance effect becomes pronounced, leading to the radiation pressure exhibiting  a well-known ``ripple structure''  \cite{ashkin1977observation, mishchenko2002scattering} as $R /\lambda_0$ varies.
The Mie resonance effect explains the ``ripple'' collection efficiency for $z$-axis depicted in \figref{fig:detection_scheme}(e,f).
Because both experimental configurations in \figref{fig:detection_scheme} exhibit sufficient information collection efficiencies to achieve three-dimensional quantum ground state cooling, we then focus on the mode-matching loss in the interferometric progress.

\subsection{Scattering field and mode-matching}
\begin{figure*}
  \centering
  \includegraphics[width=\linewidth]{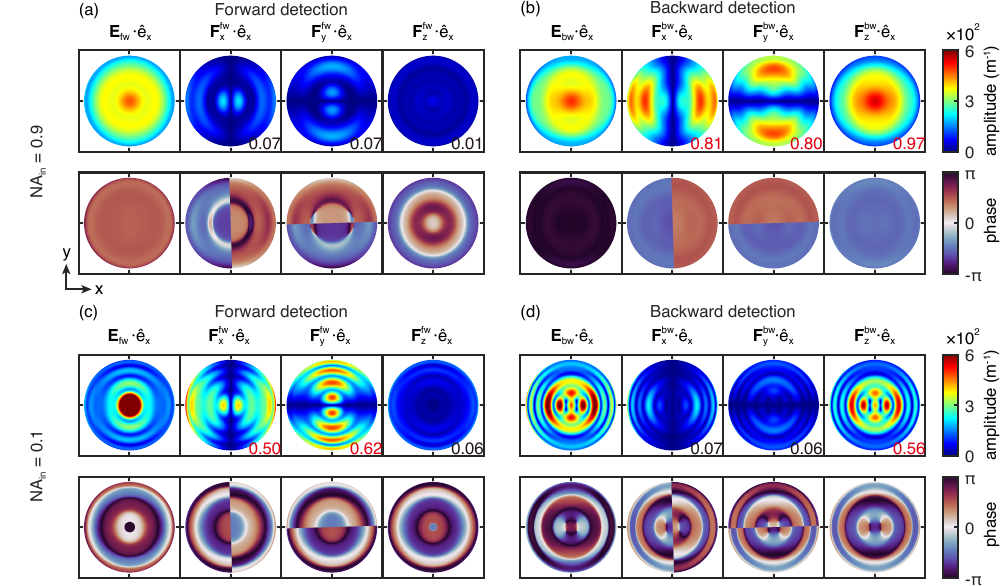}
  \caption{Forward scattering field $\mathbf{E}_{\mathrm{fw}}$ (left) and backward scattering field $\mathbf{E}_{\mathrm{bw}}$ (right) for $R / \lambda_0 = 2$ with an incident beam of numerical aperture $\mathrm{NA}_{\mathrm{in}} = 0.9$ (up) or $\mathrm{NA}_{\mathrm{in}} = 0.1$ (down) as well as corresponding normalized IRF $\mathbf{F}_{\mu}$.
  Information collection efficiencies $\eta_{\mu}^{\mathrm{c}}$ about $x$-polarizations of IRFs $\mathbf{F}_{\mu} \cdot \mathbf{\hat{e}}_{x}$ are shown in each panel, respectively.}
  \label{fig:Field4}
\end{figure*}

For the four detection schemes of the two configurations in \figref{fig:detection_scheme}, the distributions of the scattering fields on the back focal plane (BFP) as well as corresponding IRF are simulated and presented in \figref{fig:Field4}.
The simulations pertain to a Mie particle with $R / \lambda_0 = 2$.
%

In the forward detection, the forward scattering light $\mathbf{E}_{\mathrm{fw}}$ is interfered with the IRF $\mathbf{E}_{\mu}$ (see \figref{fig:Field4}(a)) and the position signal is extracted by balanced photodetector.
For a Rayleigh particle, $\mathbf{E}_{\mu} \ll \mathbf{E}_{\mathrm{fw}} \approx  \mathbf{E}_{\mathrm{in}}$ and the Gouy phase shift \cite{novotny2012principles,tebbenjohanns2019optimal} introduces a $\pi /2$ phase difference between the incident field $\mathbf{E}_{\mathrm{in}}$ and the scattered field $\mathbf{E}_{\mathrm{s}}$.
We prefer to state that the incident field $\mathbf{E}_{\mathrm{in}}$ and the IRF $\mathbf{E}_{\mu}$ are in-phase or antiphase, according to the definition of \eqnref{eq:def_IRF}.
Consequently, the incident field effectively serves as a reference field in the forward detection for a Rayleigh particle.

However, the incident field $\mathbf{E}_{\mathrm{in}}$ does not consistently serve as an effective reference field in the forward detection as $R / \lambda_0$ varies beyond the point-dipole approximation.
On the one hand, the IRF $\mathbf{E}_{\mu}$ varies with $R / \lambda_0$, causing a phase difference between the incident field $\mathbf{E}_{\mathrm{in}}$ and IRF $\mathbf{E}_{\mu}$.
On the other hand, the magnitude of the scattered field $\mathbf{E}_{\mathrm{s}}$ becomes comparable with the incident field $\mathbf{E}_{\mathrm{in}}$.
As a result, the elastically scattered field $\mathbf{E}_{\mathrm{s}}(\mathbf{r};0)$ (the first term in \eqnref{eq:expand_Es}) only increases the shot noise (denominator in \eqnref{eq:etam}) and does not contribute to the extraction of information.

Certainly, we can introduce a strong local oscillator (LO) serving as an ideal reference field to perform homodyne detection.
However, this is limited by two aspects:
(i) The LO served as a reference field whose power is needed to much larger than the total scattering field, i.e. $\mathbf{E}_{\mathrm{LO}} \gg |\mathbf{E}_{\mathrm{in}} + \mathbf{E}_{\mathrm{s}}|$.
The power threshold of the photodetector thus will limit the power of LO.
(ii) The distribution of the IRF in the forward detection changes with $R / \lambda_0$, which makes it difficult to confirm the distribution of the IRF and select the LO.
Actually, the IRF defined as the differential of the scattering field about the displacement of the particle, shown in \eqnref{eq:def_IRF} is not the directly measurable field.
The accuracy of IRF calculation depends on the accuracy of experimental parameters.

In the backward detection, the power of the backward scattering light less than one-tenth of the incident light, thereby relaxing the power threshold requirement for the photodetector.
For the configuration with $\mathrm{NA}_{\mathrm{in}} = 0.1$, the distribution of the IRF exhibits complexity (\figref{fig:Field4}d) and varies with $R / \lambda_0$, shown in \figref{fig:FWvsBW}.
Thus, it is challenging for applying an appropriate reference field in this configuration to get high mode-matching efficiency.
In contrast, for the configuration with $\mathrm{NA} _{\mathrm{in}} = 0.9$ (\figref{fig:Field4}b), the distribution of IRF corresponding to high collection detection efficiency is simple and consistent as $R / \lambda_0$ varies.
Thus, confirming the distribution of LO within this configuration is easily achievable, paving the way for get high mode matching efficiency.
\section{Scheme for Mie particle ground-state cooling}
\begin{figure}
  \centering
  \includegraphics[scale=1]{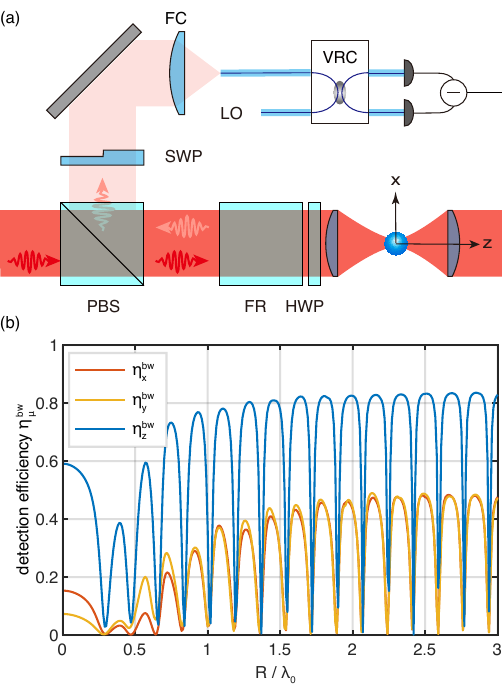}
  \caption{(a) Backward detection scheme of $\mathrm{NA}_{\mathrm{in}} = 0.9$.
  The backward scattering light is collected, modulated and coupled to perform a fiber-based balanced homodyne detection.
  HWP, half-wave plate; FR, Faraday rotator; PBS, polarizing beam-splitter; SWP, split-waveplate; FC, fiber collimator; LO, local oscillator; VRC, variable ratio coupler.
  (b) Total detection efficiency $\eta_{\mu}^{\mathrm{bw}}$ of the backward detection scheme as a function of $R / \lambda_0$ for $\mu = x, y, z$-axis.
  }
  \label{fig:BWNA09}
\end{figure}

Based on the theory and result above, we here present a comprehensive measurement scheme in which the total position detection efficiencies $\eta_{\mu}$ are sufficiently high to enable achieving ground-state cooling of three-dimensional center-of-mass motion for dielectric spheres with masses ranging from $10^{9}$ to $10^{14}$ atomic mass units.
It used the configuration of backward detection with high NA above, together with a fiber-based balanced homodyne detection as shown in \figref{fig:BWNA09}(a).

With this scheme, the total detection efficiencies $\eta_{\mu}$ within the backward detection scheme of $\mathrm{NA}_{\mathrm{in}} = 0.9$ are illustrated in \figref{fig:BWNA09}(b).
The backward detection with high NA offers several advantages:
(i) the exceptionally high collection efficiency of $\mathrm{NA}_{\mathrm{in}} = 0.9$ for specific $R / \lambda_0$, as shown in \figref{fig:detection_scheme}(b),
(ii) the consistent and simple field distribution about different $R / \lambda_0$
and (iii) the low backward scattering power.
The last two advantageous guarantee high mode-matching efficiency just with a simple optical field modulation.

After collecting the backward scattering field, a LO is introduced to match the IRF $\mathbf{E}_{\mu}$ rather than the scattered field $\mathbf{E}_{\mathrm{s}}$ to extract and amplitude the position signal.
Here, we consider a fiber-based confocal detection \cite{magrini2021realtime} for two reasons.
First, it allows easy, consistent and efficient mode matching of the local oscillator to the information light for different $R / \lambda_0$.
Second, confocal filtering by the fiber allows to suppress the stray field including the elastically scattered field $\mathbf{E}_{\mathrm{s}}$ and stray reflections field.
For longitudinal detection, the longitudinal IRF $\mathbf{E}_{z}$ is directly coupled into a single-mode fiber (SMF).
As a contrast, for lateral detection, the scattering field $\mathbf{E}_{\mathrm{s}}$ and lateral IRF $\mathbf{E}_{x(y)}$ must be modulated before coupling the SMF.
This modulation is achieved by using a split-waveplate (SWP) or a mode-converter and represented by $\mathbf{E}^{\mathrm{mod}}(X,Y) = t_{x,(y)}(X, Y) \mathbf{E}(X,Y)$, where $t_{x(y)}(X, Y)$ is defined as:
%
\begin{equation}\label{eq:t_SWP}
  \begin{aligned}
    t_{x}(X) & = \left\{\begin{matrix}
                          1,  & \mathrm{for}~X>0 \\
                          -1, & \mathrm{for}~X<0
                        \end{matrix} \right., \\
    t_{y}(Y) & = \left\{\begin{matrix}
                          1,  & \mathrm{for}~Y>0 \\
                          -1, & \mathrm{for}~Y<0
                        \end{matrix} \right.. \\
  \end{aligned}
\end{equation}
This modulation changes the symmetry of both the backward scattering field $\mathbf{E}_{\mathrm{bw}}$ and lateral IRF $\mathbf{E}_{x,y}$, thereby facilitating the coupling of the IRF into the SMF and filtering the scattering field without information.
This technique is referred to as structured light detection \cite{madsen2021ultrafast,li2023structuredlight}.

The magnification $(M = f_3 / f)$ of the confocal detection is selected to be $9.2$ in this paper, where $f_{3}$ and $f$ denote the focal length of the fiber condenser and the objective lens, respectively.
This magnification ensures effective coupling of the IRF into the SMF.
Further discussion on the effect of magnification on the mode-matching efficiency of the backward detection is provided in \appref{app:confocal_detection}.

\section{Conclusion}

In conclusion, we have theoretically investigated the problem of optimal position detection of a Mie particle.
We introduced the IRF to characterize the information carried by the scattered field about the center of mass position of the particle.
The characteristics of Mie scattering provides the sufficiently high information collection efficiency in the backward or forward detection.
However, particular attention must be given to mode-match efficiency to effectively extract information about the particle's center of mass position.
In the forward detection configuration, the rapidly changing IRF distribution and intense forward scattering light poses limitations on the mode-matching in the homodyne detection.
Conversely, in the backward detection configuration, strategic selection of the numerical aperture of the lens and the filling factor of the incident beam can provide sufficient position detection efficiency with simple light field modulation.
Based on these results, we also proposed an experimental scheme that can achieve three-dimensional center-of-mass ground-state cooling of Mie particles with masses of $10^{14}$ atomic mass units.

Our theory for optimal measurement can be extended from the translational motion of micrometer-scale dielectric spherical particles to various shapes of particles and different degrees of freedom, including libration \cite{vanderlaan2021subkelvin, zielinska2023controlling, kamba2023nanoscale, gao2024feedback} and internal acoustic vibrations \cite{jesipe2016hamiltonian, zoubi2016optomechanical}.
The calculation of the scattered field relies on the acquisition of the $T$-matrix of the particle. Therefore, precise characterization of the particle is crucial, including its mass \cite{zheng2020robust, tian2022medium, tian2023temperaturefree} and anisotropy \cite{rademacher2022measurement}, as well as power spectral analysis \cite{liu2024nano}.
This enables shot-noise-limited detection across all degrees of freedom for larger particles, including Mie particles, and facilitates various applications based on macroscopic quantum states.
Specifically, a high-frequency gravitational-wave detector based on a levitated sensor detector (LSD) using optically levitated dielectric discs \cite{aggarwal2022searching, arvanitaki2013detecting} could benefit from our methods. Unlike laser interferometer detectors like ALIGO and VIRGO, which are limited by photon shot noise, LSD systems are constrained by the center-of-mass (c.m.) motion temperature of the levitated particles, which could be improved using our techniques.
Furthermore, by implementing vacuum squeezing injection \cite{gonzalez-ballestero2023suppressing}, displacement sensitivities beyond the standard quantum limit could be achieved. Consequently, levitating micrometer-sized particles represent a promising platform for exploring new physical phenomena, such as dark matter \cite{afek2021limits, afek2022coherent, carney2023searches, kawasaki2020high, monteiro2020search, moore2021searching, priel2022dipole, priel2022search, rademacher2020quantum}.
Lastly, Mie particles exhibit fast recoil heating rates beyond the dipole-point approximation, comparable to and sometimes exceeding mechanical frequencies, enabling the exploration of the strong quantum optomechanical regime \cite{das2023instabilities, dare2023linear, delosriossommer2021strong, forn-diaz2019ultrastrong, friskkockum2019ultrastrong, monteiro2013dynamics} in free space without needing optical resonators.
Of course, stronger interaction between light and the particle increases the difficulty of feedback cooling \cite{gieseler2012subkelvin, tebbenjohanns2019cold, zheng2019cooling, vijayan2023scalable}.

\begin{acknowledgments}
  This work was supported by the National Natural Science Foundation of China (Grant No. 12104438, No. 62225506 and No. 12204140), the CAS Project for Young Scientists in Basic Research (Grant No. YSBR-049), and the Fundamental Research Funds for the Central Universities.
\end{acknowledgments}

%

\clearpage

\onecolumngrid
\appendix

\section{Calculation in forward and backward detection}\label{app:Field_cal}
In this appendix, we calculate the fields in realistically interferometric measurements.
Firstly, we calculate the total forward and backward scattering field as well as the corresponding IRF on the BFP of the collection lens.

We consider an $x$-polarized Gaussian beam field $\mathbf{E}_{\mathrm{in}}$ propagating along the $z$-axis, tightly focused by an objective lens.
A silica dielectric sphere is assumed to be trapped at the focus.
The origin of the coordinate system is defined by the focus where the particle makes a tiny motion.
Then the scattered field $\mathbf{E}_{\mathrm{s}}$ can be calculated by $T$-matrix method \cite{jones2015optical} and its open source package $ots$.
The total scattering field $\mathbf{E}_{\mathrm{tot}} = \mathbf{E}_{\mathrm{in}} + \mathbf{E}_{\mathrm{s}}$ constitutes the solution of Mie problem.
Consider mapping the field at point $\mathbf{r} = (f, \theta, \phi)$ onto the focal plane plane $(X, Y)$ of the collection lens, then the forward scattered field $\mathbf{E}_{\mathrm{fw}}$ is
\begin{equation}\label{eq:E_fw}
  \mathbf{E}_{\mathrm{fw}}(X, Y) = \frac{1}{\sqrt{\cos \theta}} \{ [\mathbf{E}_{\mathrm{s}}(\mathbf{r}) \cdot \mathbf{\hat{e}}_{\theta}] \mathbf{\hat{e}}_{\rho} + [\mathbf{E}_{\mathrm{s}}(\mathbf{r}) \cdot \mathbf{\hat{e}}_{\phi}] \mathbf{\hat{e}}_{\phi} \} + \mathrm{e}^{i \pi} \mathbf{E}_{\mathrm{in}}(X, Y)
\end{equation}
where the unit vectors $\mathbf{\hat{e}}_{\theta}, \mathbf{\hat{e}}_{\phi}, \mathbf{\hat{e}}_{\rho}$ can be expressed in terms of the Cartesian unit vectors $\mathbf{\hat{e}}_{x}, \mathbf{\hat{e}}_{y}, \mathbf{\hat{e}}_{z}$ as:
\begin{equation}\label{eq:unit_vectors}
  \begin{aligned}
    \mathbf{\hat{e}}_{\rho}   & = \cos \phi~\mathbf{\hat{e}}_{x} + \sin \phi~\mathbf{\hat{e}}_{y},                                                            \\
    \mathbf{\hat{e}}_{\phi}   & = -\sin \phi~\mathbf{\hat{e}}_{x} + \cos \phi~\mathbf{\hat{e}}_{y},                                                           \\
    \mathbf{\hat{e}}_{\theta} & = \cos \theta \cos \phi~\mathbf{\hat{e}}_{x} + \cos \theta \sin \phi~\mathbf{\hat{e}}_{y} + \sin \theta~\mathbf{\hat{e}}_{z}, \\
  \end{aligned}
\end{equation}
The backward scattering field $\mathbf{E}_{\mathrm{fw}}$ is equivalent to the backward scattered field $\mathbf{E}_{\mathrm{s}}$ and can also be calculated, just by replacing $\theta$ to $\pi - \theta$, namely,
\begin{equation}\label{eq:E_bs}
  \mathbf{E}_{\mathrm{bw}}(X, Y) = \frac{1}{\sqrt{|\cos \theta|}} \{ [-\mathbf{E}_{\mathrm{s}}(\mathbf{r}) \cdot \mathbf{\hat{e}}_{\theta}] \mathbf{\hat{e}}_{\rho} + [\mathbf{E}_{\mathrm{s}}(\mathbf{r}) \cdot \mathbf{\hat{e}}_{\phi}] \mathbf{\hat{e}}_{\phi} \}.
\end{equation}
The IRF $\mathbf{E}_{\mu}$ of backward and forward detection follow the same calculation based on the definition of \eqnref{eq:def_IRF}.
\begin{figure}
  \centering
  \includegraphics[width=\linewidth]{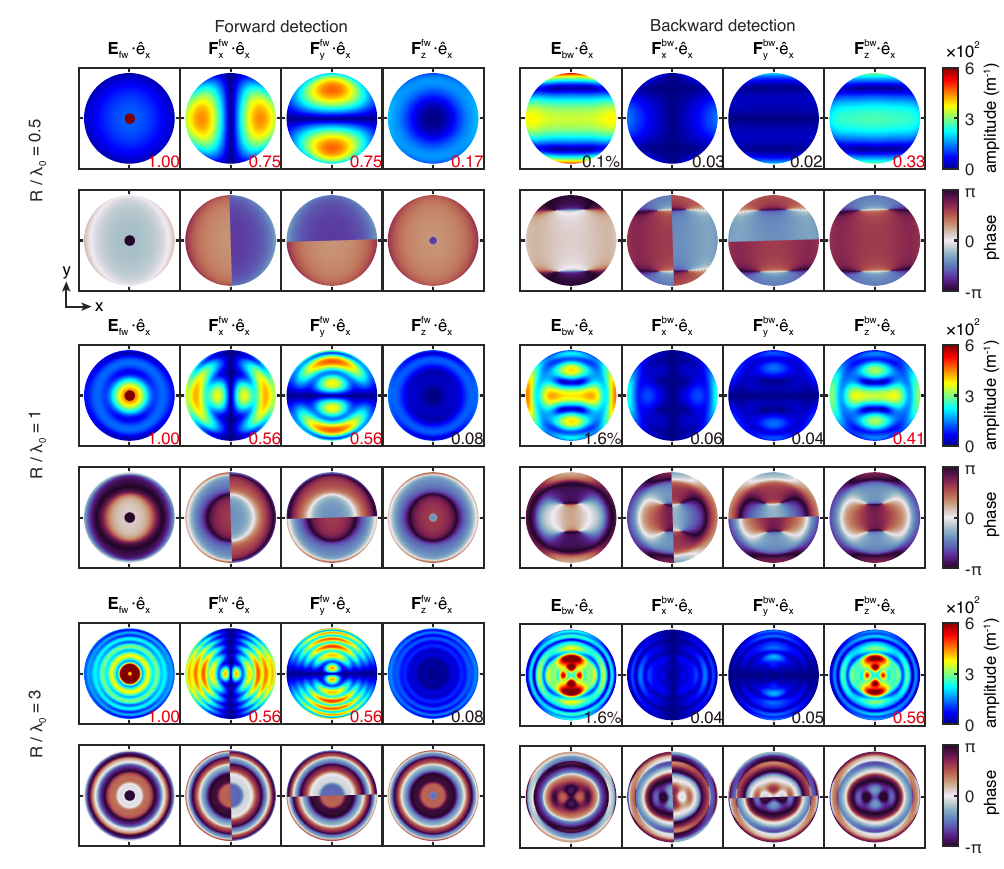}
  \caption{Forward scattering field $\mathbf{E}_{\mathrm{fw}}$ (left) and backward scattering field $\mathbf{E}_{\mathrm{bw}}$ (right) for an incident beam of numerical aperture $\mathrm{NA}_{\mathrm{in}} = 0.1$ as well as corresponding normalized IRF $\mathbf{F}_{\mu}$ as functions of the silica dielectric sphere's size parameter $R / \lambda_0$.
  We show the mainly $x$-polarizations of the fields.}
  \label{fig:FWvsBW}
\end{figure}

In the case of $\mathrm{NA}_{\mathrm{in}} = 0.1$, shown in \figref{fig:FWvsBW}, the lateral IRF $\mathbf{F}_{x(y)}^{\mathrm{fw}}$ provides the high information collection efficiency in the forward detection of $R /\lambda_0 = 0.5$.
The distribution of the IRF makes it easy and efficient for interferometric measurement, that is shading the forward scattering field $\mathbf{E}_{\mathrm{fw}}$ and applying an mode-matching local oscillator for homodyne detection.
However, the fields become complex as $R / \lambda_0$ increases.
This makes it difficult for applying an ideal local oscillator, as we have discussed in the main text.

\begin{figure}
  \centering
  \includegraphics[width=\linewidth]{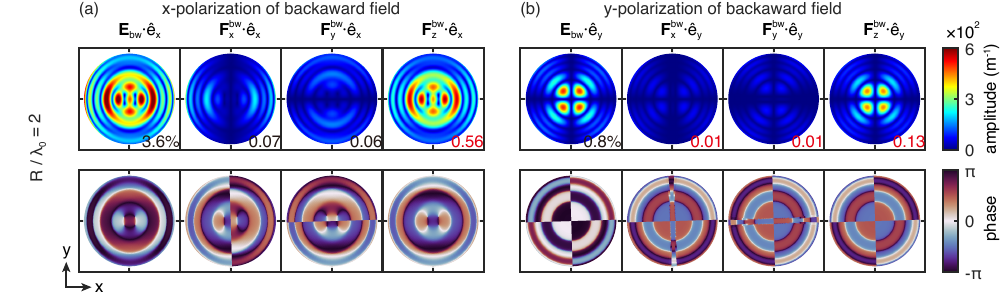}
  \caption{Compare (a) $x$-polarization with (b) $y$-polarization of the backward scattering field $\mathbf{E}_{\mathrm{bw}}$ for an incident beam of numerical aperture $\mathrm{NA}_{\mathrm{in}} = 0.1$ as well as corresponding normalized IRF $\mathbf{F}_{\mu}$.
  It shows that the $y$-polarization cannot be ignored in the backward detection of $\mathrm{NA}_{\mathrm{in}} = 0.1$.}
  \label{fig:BWxy}
\end{figure}

Furthermore, it's worth noting that the $y$-polarization of the fields cannot be ignored in the backward detection of $\mathrm{NA}_{\mathrm{in}} = 0.1$.
This means that the information carried the $y$-polarization of the backward IRF cannot be ignored, as illustrated in \figref{fig:BWxy}.

\section{Fiber-based confocal detection}\label{app:confocal_detection}
In this section, we show calculation about the fiber-based confocal detection of numerical aperture $\mathrm{NA}_{\mathrm{in}} = 0.9$ of the incident beam.
The backward scattering field is collected and coupled into a single-mode fiber and it must be modulated by the split-waveplate (SWP) for lateral detection comparing with longitudinal detection, as we have discussed in the main text.

The arbitrary field on the focal plane $(X_{\mathrm{f}}, Y_{\mathrm{f}})$ of the fiber condenser is calculated by using the angular spectrum integration \cite{novotny2012principles}, expressed as
\begin{equation}\label{eq:ASR}
  \mathbf{E}(\rho, \varphi, z) = - \frac{\mathrm{i} k f \mathrm{e}^{-\mathrm{i} k f}}{2\pi} \int\limits_{0}^{\theta_{\max}} \int\limits_{0}^{2\pi} \mathbf{E}_{\infty}(\theta, \phi) \mathrm{e}^{\mathrm{i} k z \cos \theta} \mathrm{e}^{\mathrm{i} k \rho \sin \theta \cos(\phi-\varphi)}\sin \theta \mathrm{d}\phi \mathrm{d}\theta
\end{equation}
where the cylindrical coordinate $(\rho, \varphi, z)$ corresponding to $(X_{\mathrm{f}}, Y_{\mathrm{f}})$ and the far-field $\mathbf{E}_{\infty}$ can be expressed as
\begin{equation}\label{eq:E_far}
  \mathbf{E}_{\infty}(\theta, \phi) = [(\mathbf{E}(X, Y) \cdot \mathbf{\hat{e}}_{\phi})\mathbf{\hat{e}}_{\phi} + (\mathbf{E}(X, Y) \cdot \mathbf{\hat{e}}_{\rho})\mathbf{\hat{e}}_{\theta}] \sqrt{\cos \theta}.
\end{equation}
where $\mathbf{\hat{e}}_{\theta} = \cos \theta \cos \phi~\mathbf{\hat{e}}_{x} + \cos \theta \sin \phi~\mathbf{\hat{e}}_{y} - \sin \theta~\mathbf{\hat{e}}_{z}$ is different from $\mathbf{\hat{e}}_{\theta}$ in \eqnref{eq:unit_vectors}.
The focused fields on the fiber plane are calculated just by considering $\mathbf{E}$ as $\mathbf{E}_{\mathrm{bw}}$ or $\mathbf{E}_{\mu}$.

The LO is the basic mode of the SMF, the linear polarized $\mathrm{LP}_{01}$ mode \cite{vamivakas2007phasesensitive}, denoted as
\begin{equation}\label{eq:LP_01}
  \mathbf{E}_{01}(\mathbf{r}, t) = \left\{\begin{matrix}
    \displaystyle N J_0\left( \frac{u r}{a_{\mathrm{co}}} \right) \mathrm{e}^{\mathrm{i} \beta z} \mathbf{\hat{e}}_{x},                       & r \le a_{\mathrm{co}} \\
    \displaystyle N \frac{J_0(u)}{K_0(w)} K_0\left( \frac{w r}{a_{\mathrm{co}}} \right) \mathrm{e}^{\mathrm{i} \beta z} \mathbf{\hat{e}}_{x}, & r \ge a_{\mathrm{co}} \\
  \end{matrix} \right.
\end{equation}
where $N$ is a normalization constant, $u = a_{\mathrm{co}}\sqrt{n^{2}_{\mathrm{co}} k_0^{2} - \beta^{2}}$ and $w = a_{\mathrm{co}} \sqrt{\beta^{2} - n^{2}_{\mathrm{cl}} k_0^{2}}$ are the lateral wavenumbers, $V^{2} = u^{2} + w^{2} = a k_0 \sqrt{n^{2}_{\mathrm{co}} - n^{2}_{\mathrm{cl}}}$ is the fiber $V$-parameter. The $J_{l}$ and $K_{l}$ represent the Bessel functions of the first and second kind, respectively, with order $l$. The propagation constant $\beta$ needs to be solved by solving the eigenequation of $\mathrm{LP}_{01}$ mode, expressed as
\begin{equation}\label{eq:solve_beta}
  \frac{J_0(u)}{u J_1(u)} - \frac{K_0(w)}{w K_1(w)} = 0.
\end{equation}
\begin{figure}
  \centering
  \includegraphics[width=\linewidth]{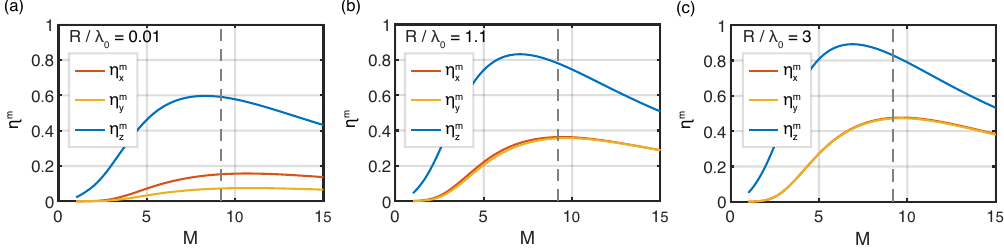}
  \caption{Mode-matching efficiency $\eta_{\mu}^{m}$ as a function of magnification $M$ of the fiber-based confocal system and the particle radius $R$ for all three axes $\mu = x, y, z$.
    (a-c) Mode-matching efficiency for $R /\lambda_0 = 0.01, 1.1, 3$, respectively.
    The gray dotted line $M = 9.2$ shows our choosing magnification in this paper, not far from the optimal value for different $R /\lambda_0$ and three axes $\mu = x, y, z$.
    $M = 10$ corresponds to the maximum mode-matching efficiency of lateral position detection.}
  \label{fig:etamVSM}
\end{figure}

The mode-matching efficiency of the SLD can be calculated by using \eqnref{eq:etam} in the main text.
Then the mode-matching efficiency as a function of magnification $M = f_3 / f_{\mathrm{c}}$, the focal length ration of the fiber condenser $f_3$ to the collection lens $f_{\mathrm{c}}$, is shown in \figref{fig:etamVSM}.

Thus, a magnification value of $M = 9.2$ is selected for the calculation of confocal efficiency (see \figref{fig:BWNA09}).
This choice corresponds to focal length of commercial objective lens $f_{\mathrm{c}} = \SI{2}{\mm}$ and fiber condenser $f_3 = \SI{18.4}{\mm}$.
The custom magnification and the appropriate $z$-position of the SMF can further improve the coupling efficiency.

\end{document}